\documentclass[aps,pr,twocolumn,showpacs,superscriptaddress,groupedaddress,nofootinbib]{revtex4}  

\usepackage{graphicx}  
\usepackage{dcolumn}   
\usepackage{bm}        
\usepackage{amssymb}  
\usepackage{amssymb}
\usepackage{amsxtra}
\usepackage{euscript} 
\usepackage{amsthm}
\usepackage{color}

\begin{document}

\title{Functional Renormalization Group analysis of rank 3 tensorial group field theory: \\
The full quartic invariant truncation}

\author{Joseph Ben Geloun}
\email{bengeloun@lipn.univ-paris13.fr}
\affiliation{Laboratoire d'Informatique de Paris Nord UMR CNRS 7030
Universit\'e Paris 13, 99, avenue J.-B. Clement, 93430 Villetaneuse, France}
\affiliation{International Chair in Mathematical Physics and Applications
ICMPA -- UNESCO Chair, 072 B.P. 50 Cotonou, Benin}

\author{Tim A. Koslowski}
\email{koslowski@nucleares.unam.mx}
\affiliation{Instituto  de  Ciencias  Nucleaes,  Universidad  Nacional  Aut\'onoma  de  M\'exico,
Apdo.   Postal  70-543,  04510  Ciudad  de  M\'exico,  D.F.,  M\'exico}

\author{Daniele Oriti}
\email{daniele.oriti@aei.mpg.de}
\affiliation{Max Planck Institute for Gravitational Physics (Albert Einstein Institute),
Am M\"uhlenberg 1, D-14476 Potsdam-Golm, Germany, EU}
\affiliation{II Institute for Theoretical Physics, University of Hamburg,  Luruper Chaussee 149, 22761 Hamburg, Germany, EU}

\author{Antonio D. Pereira}
\email{a.pereira@thphys.uni-heidelberg.de}
\affiliation{Institute for Theoretical Physics, University of Heidelberg, Philosophenweg 12, 69120 Heidelberg, Germany}

\begin{abstract}
\noindent In this paper we consider the complete momentum-independent quartic order truncation for the effective average action of a real Abelian rank 3 tensorial group field theory. This complete truncation includes non-melonic as well as double-trace interactions. In the usual functional renormalization group perspective, the inclusion of more operators that belong to the underlying theory space corresponds to an improvement of the truncation of the effective average action. We show that the inclusion of non-melonic and double-trace operators in the truncation brings subtleties. In particular, we discuss the assignment of scaling dimensions to the non-melonic sector and how the inclusion of double-trace operators considerably changes the results for critical exponents when they are not included. We argue that this is not a particular problem of the present model by comparing the results with a pure tensor model. We discuss how these issues should be investigated in future work.
\end{abstract}

\maketitle

\section{Introduction}
Attempts to quantise the gravitational field started soon after the very formulation of General Relativity (the current classical theory of the gravitational field) and of quantum field theory, the modern framework for describing fundamental interactions \cite{Kiefer:2004gr}. While suggestions that a more radical departure from standard spacetime-based physics were also voiced, the first strategy was to apply to the gravitational field the same techniques that have been successfully applied to other interactions, turning General Relativity into a quantum field theory. In particular, the perturbative quantization of the metric field around flat spacetime promised to suffice for both phenomenological and theoretical purposes. The proof of perturbative non-renormalizability of such quantum field theory \cite{tHooft:1974toh,Christensen:1979iy,Goroff:1985th} showed that, while possibly phenomenologically adequate in some limited regime \cite{Donoghue:1993eb}, it could not be a fundamental description of the gravitational field, of spacetime and geometry at the quantum level. The reactions to this stumbling block have been diverse. Many took it to imply that, while a field-theoretic approach aiming at a straightforward quantization fo General Relativity was still viable, it was crucial to work directly at the non-perturbative level, using a different set of quantization tools. This is the case of the canonical quantization ala Dirac leading to loop quantum gravity \cite{Thiemann:2007zz,Ashtekar:2004eh,Rovelli:2004tv}, of the lattice path integrals on which quantum Regge calculus and dynamical triangulations \cite{Ambjorn:2012jv,Laiho:2016nlp} are based (together with spin foam models \cite{Perez:2012wv,Rovelli:2014ssa}, a covariant counterpart of loop quantum gravity), and of the asymptotic safety scenario for the quantum gravitational field based on functional renormalization group equations \cite{Reuter:2012id,Eichhorn:2017egq}. String theory, while initially formulated within the same perturbative viewpoint, quickly moved beyond it too, but also pointed towards a radical generalization of local quantum field theory identifying in the need for such generalization the real culprit of perturbative quantum gravity. It also provided hints that we may have to move beyond spacetime physics as well, and that spacetime and geometry may have to be considered as emergent notions \cite{Oriti:2013jga,Oriti:2017twl,Seiberg:2006wf}. The idea of an emergent spacetime, actually, has become quite widespread in recent years and started playing an important role in several quantum gravity formalisms, moving further beyond the idea of a quantum field theory formulation of quantum gravity as they move beyond any reliance on spacetime itself.

Tensorial group field theories\footnote{A clarification on the nomenclature is in order. Group field theories, tensor models, tensor field theories and tensorial group field theories are used in the literature for models of the type we deal with in this paper, and that we are going to specify better in the following. The use of one or the other label may depend on the context, the aim of the work or the emphasis on one or the other feature of the formalism. The \lq tensor\rq~ label emphasizes the fact that the basic dynamical variable is a tensor, and that its transformation properties are crucially used in determining the relevant theory space. The \lq field\rq~ label emphasizes the existence of non trivial propagators, softly breaking the invariance of the interactions, or the fact that the model possesses an infinity of degrees of freedom, with the basic dynamical variable being a field on some domain space. The label \lq group\rq~emphasizes that such domain space is usually a Lie group manifold or the fact that the ensuing group structures are heavily used to define the model’s action and/or its symmetries. In this paper we use the most general label, also because all the corresponding features, tensorial nature, group structure and field-theoretic nature, are present and important.} (TGFTs) \cite{BenGeloun:2012yk,Geloun:2013saa,Geloun:2015lta,Geloun:2012bz,BenGeloun:2011rc,BenGeloun:2012pu,Geloun:2014ema,Samary:2014tja,Samary:2012bw,Samary:2014oya} represent both a confirmation and a contradiction of this line of developments. They are a proposal for {\it a quantum field theory of spacetime} (and thus of quantum gravity), which is then necessarily defined without relying on any (pre-existing) spacetime and in which the latter has to be emergent rather than fundamental. They are a class of quantum field theories where, at the same time, one tries to apply standard field-theoretic techniques familiar from the theory of fundamental interactions, and is forced to modify, adapt and sometimes drop several of them, due to the peculiar spacetime-free setting. They incarnate a very different perspective on quantum gravity, with respect to other approaches mentioned above, and based on the wish for a straightforward quantization of General Relativity, while managing to incorporate several of their defining structures and being historically a further development of them. With details depending on the specific model being considered, TGFTs can be seen to define a generating functional for a theory of random lattices \cite{Rivasseau:2011hm,Rivasseau:2012yp}, like in dynamical triangulations, a second quantized reformulation of canonical loop quantum gravity \cite{Oriti:2013aqa}, a complete definition of spin foam models \cite{Oriti:2014yla}, and a completion of lattice gravity path integrals via their embedding in a quantum field theory framework \cite{{Baratin:2011hp}}. In a way, TGFTs bring together but also move beyond these related quantum gravity formalisms, and show that a quantum field theory formulation of quantum gravity and geometry can in principle be achieved, if we are willing to move beyond spacetime and usual field theories, in order to do so.  

The quantum field theory setting of TGFTs allows to use key tools from QFT to tackle outstanding issues of quantum gravity approaches. The quantum consistency and the construction (and quantization) ambiguities of TGFT models can be constrained by the request for (perturbative) renormalizability of the same, while the crucial problem of defining the continuum limit and thus the full quantum partition function of the same models (including all their infinite degrees of freedom) becomes the problem of defining their non-perturbative renormalization group flow and phase diagram.  

TGFT renormalization has become in fact a key area of rapid developments and many results (see \textit{e.g.} \cite{BenGeloun:2012yk,Geloun:2013saa,Geloun:2015lta,Geloun:2012bz,BenGeloun:2011rc,BenGeloun:2012pu,Geloun:2014ema,Samary:2014tja,Samary:2012bw,Samary:2014oya,Carrozza:2012uv,Carrozza:2013wda,Carrozza:2014rya,Carrozza:2014rba,Lahoche:2015tqa,Lahoche:2015ola} as well as \cite{Carrozza:2016vsq} for a review). These results include the proof of perturbative renormalizability of a variety of models (abelian and non-abelian, in different dimensions, with and without gauge symmetries) and the establishment of their asymptotic freedom or safety; for more involved models (those more closely related to loop quantum gravity, spin foam models and 4d simplicial gravity path integrals), detailed studies of their divergences and radiative corrections; results on constructive renormalization; the application of functional renormalization group (FRG) techniques to establish the RG flow of various models, and useful indications about their phase diagram. 

Many open issues remain, of course. Among them, the key one seems to us to be the uncertainty on the relevant TGFT theory space, for the most interesting (and involved) 4d quantum gravity models but also for the simpler ones. Part of this uncertainty is due to the limits of conventional QFT wisdom when applied to TGFTs and to the lack of knowledge about key aspects of the formalism like symmetries (but see \cite{BenGeloun:2011cz,Kegeles:2015oua,Kegeles:2016wfg}) or its mathematical foundations (but see \cite{Kegeles:2017ems}). Part is due simply to the still limited experience we have accumulated with the RG flow of such models, the exact consequences of the combinatorial structure of their possible interactions, and the difficulties in guessing, therefore, the universal aspects of their phase diagrams.  

In this paper, we offer one more exploration of TGFTs and a study of their RG flow, using functional techniques, aimed exactly at understanding better their theory space and the role that different combinatorial structures in their interactions have on the same flow. We perform the FRG analysis of a real, $U(1)$ TGFT model of rank 3, without gauge invariance, in the {\it most extended} truncation of order 4 and momentum-independent interactions terms. That is, we extend the existing analysis of a similar model at the same order \cite{Benedetti:2014qsa} by including both non-melonic interaction vertices and \lq disconnected\rq~ vertices, i.e. those defined by multiple independent integrals of tensor invariant functionals of the field. Non-melonic interactions are expected to be subdominant in the large-N limit of the theory, i.e. in the deep \lq UV\rq, and for this reason they are usually left out (but see \cite{Carrozza:2017vkz}). Thus, they are excluded for reasons internal to the TGFT formalism, since they have no analogue in usual QFT. Disconnected vertices, on the other hand, could be defined also in usual QFT, where they correspond to non-local interactions and are thus excluded on physical grounds; this exclusion, though, is consistent with the RG flow in the sense that such terms are not generated by it, when initially excluded. This is not the case in TGFT, where we also lack any clear physical motivation for excluding them. However unusual and of unclear physical significance they may be (from the point of view of lattice quantum gravity as well), consistency of the formalism would suggest to include them in the action and study their effect on the RG flow. This is what we do.
As to be expected for TGFTs on compact groups \cite{Benedetti:2014qsa}, we end up with a non-autonomous system of RG equations. We limit our analysis to the UV regime of the model where, by an appropriate rescaling of the coupling constants, the system becomes autonomous and fixed points can be identified.

Our main results are going to be discussed at some length in the concluding section. Here, it suffices to say that we expose some ambiguities in the rescaling of the coupling constants, and some more hints that the 4th order truncation may not give reliable and conclusive indications of the actual RG flow of the model. Also, we find that the extension of the truncation to non-melonic diagrams does not lead, in most cases, to drastic changes in the flow, while the effects of the disconnected diagrams can actually be more profound, calling for a careful consideration of their role in TGFTs.

\section{Functional Renormalization Group for TGFTs}

In order to probe non-trivial fixed points in the renormalization group flow of TGFTs we will make use of the functional renormalization group (FRG), see \cite{Berges:2000ew,Pawlowski:2005xe,Rosten:2010vm} for general reviews. The FRG equation is a flow equation for the effective average action
\begin{equation}
 \Gamma_N[\phi]:=\sup_J\left(\langle J,\phi\rangle-W_N[J]\right)-\frac{1}{2}\langle\phi,R_N\,\phi\rangle,
\end{equation}
where $W_N[J]$ denotes the generating functional of connected Feynman diagrams and $\langle J,\phi\rangle$ denotes the canonical pairing between the field $\phi$ and its dual, and where $\phi$ can be considered as the vacuum expectation value of the quantum field in the presence of a source $J$. This is calculated using the IR-modified generating functional
\begin{equation}
  e^{W_N[J]}:=\int[d\varphi]_{N^\prime}\exp\left(-S[\varphi]-\frac 1 2 \langle\varphi,R_N\,\varphi\rangle+\langle J,\varphi\rangle\right),
\end{equation}
where one adds a scale dependent mass term - represented by $\langle\varphi,R_N\,\varphi\rangle$ - which gives a mass of order $N$ to modes in the IR of $N$ and which essentially vanishes for modes in the UV of $N$ (we call modes in the IR/UV of $N$ as those with ``momenta" smaller/greater than $N$). This scale dependent mass term serves as an IR-suppression term (i.e. a large Boltzmann factor) for modes in the IR of $N$. Note that we assume the functional integral to be UV-regulated at a scale $N^\prime$. It follows that the effective average action satisfies the FRG equation
\begin{equation}
\partial_t\Gamma_N = \frac{1}{2}\mathrm{Tr}\left[\frac{\partial_t R_N}{\Gamma^{(2)}_N+R_N}\right]\,,
\label{frg1}
\end{equation}
where we introduced the shorthand $t=\ln(N)$. Moreover, it follows from the properties of the IR-suppression term $\langle \phi ,R_N\phi\rangle$ (see \cite{Berges:2000ew,Pawlowski:2005xe,Rosten:2010vm}) that the limit $N\to 0$ of $\Gamma_N[\phi]$ coincides with the standard effective action $\Gamma[\phi]$, while the limit $N\to\infty$ (always keeping $N<N^\prime$) yields the bare action $S_{ren}[\phi]$ with fully renormalized couplings. This interpolation property between bare and effective action allows one to use the FRG as a non-perturbative tool to investigate possible continuum limits and their universality classes. 

In particular, if one chooses an IR suppression term such that $\partial_t R_N=0$ for $N=N^\prime$, then one can freely move the UV regulator $N^\prime$ without introducing an additional $N^\prime$ dependence in the flow equation. So practically, one can forget about the $N^\prime$ dependence altogether and investigate the possible continuum limits by investigating the flow in the limit $N\to\infty$ alone, although strictly speaking it is obtained as $N\to N^\prime\to\infty$. This feature of the FRG has been used in \cite{Eichhorn:2013isa,Eichhorn:2014xaa} to investigate the continuum limits of matrix models and this setting was extended to investigate the continuum limits of tensor models in \cite{Eichhorn:2017xhy} and of tensorial (group) field theories in \cite{Benedetti:2014qsa,Geloun:2015qfa,Geloun:2016qyb,Benedetti:2015yaa,Lahoche:2016xiq,Carrozza:2016vsq,Carrozza:2016tih,Carrozza:2017vkz,Lahoche:2018vun,Geloun:2016xep}. The relation of these models to discrete gravity can be seen at the level of their Feynman amplitudes. 
The Feynman diagrams $\gamma$ of matrix and (un-)colored tensorial models are dual to simplicial complexes $\Delta(\gamma)$ of $d$-dimensional pseudo manifolds (where $d$ is the rank of the tensor). Hence, the Feynman expansion of the partition function
\begin{equation}
 e^{W}=\sum_\gamma\,A(\gamma)=\sum_{\Delta(\gamma)}\,e^{\ln(A(\gamma))},
\end{equation}
where $A(\gamma)$ denotes the amplitude of the Feynman graph $\gamma$, can be interpreted as the sum over triangulations with a Boltzmann factor given by the Feynman amplitude. 
For simple tensor models as well as for the simplest Abelian tensorial (group) field theories without gauge invariance, the Feynman amplitude is purely combinatorial and can be put in direct correspondence with the Regge action for gravity discretized on a piecewise-flat equilateral triangulation $A(\gamma)=\exp(g_{D-2}\,N_{D-2}-g_D\,N_D)$, where $N_D$ denotes the number of $D$ simplices in the dual triangulation $\Delta(\gamma)$. The resulting partition function coincides with that of (Euclidean) dynamical triangulations. Formally, a continuum limit could be taken as follows. Using e.g. the amplitude of colored tensor models 
$$A(\gamma)=e^{\ln(N)\,N_{D-2}-\left(\ln(g)-\frac{D(D-1)}{4}\ln(N)\right)\,N_D}$$
one notes that one can take the continuum limit (in which the volume $V_o$ of the $D$-simplices approaches 0) at fixed physical volume $V=V_o\,N_D$ by taking the limit $N\to\infty$ while taking the fiducial volume of the simplex $V_o\to 0$ in a way that ensures that $V$ stays fixed. This large $N$ limit corresponds to $1/G\propto g_{D-2}\to \infty$, i.e. to the limit of vanishing Newton's constant. To retain a continuum limit at finite value of $G$, one needs to scale the coupling constant $g$ in a nontrivial way with $N$ as one takes the large $N$ limit. This means that one searches for a non-trivial fixed point of the RG flow in the scale $N$ of the matrix or tensor model, which establishes the FRG as a tool to investigate possible continuum limits in tensorial models of quantum gravity \cite{Eichhorn:2013isa,Eichhorn:2014xaa}. This is the field-theoretic, TGFT counterpart of the double scaling limit of matrix models as the continuum limit of two dimensional Euclidean quantum gravity \cite{Weingarten:1982mg,David:1984tx,David:1985nj,Kazakov:1985ea,Ambjorn:1985az,Boulatov:1986mm,DiFrancesco:1993cyw}.

For more involved TGFT models (e.g. for models directly related to loop quantum gravity), the relation to discrete (quantum) gravity is more involved. Still, the general correspondence is analogous: the Feynman amplitudes of TGFT models correspond to simplicial gravity path integrals, on the triangulations dual to the model's Feynman diagrams \cite{Baratin:2011hp}. The group-theoretic data characterizing the TGFT variables (and the domain of the fundamental field) acquire the interpretation (when suitably chosen) of discrete gravity data, i.e. a discrete connection and discrete metric variables (encoding, ultimately, the dynamical edge lengths of the triangulation), which are then summed over to define the given Feynman amplitude weighted by the exponential of a richer discrete gravity action function of them, as in quantum Regge calculus. One could indeed expect that such additional discrete geometric data are needed to capture the increased complexity of gravity and geometry in going from two dimensions, where matrix models and random equilateral surfaces suffice to define (Euclidean) quantum gravity, to higher dimensions. Of course, this leads to more complicated models and to much richer theory spaces, which are not yet under control.

The more complete and detailed explorations of TGFT models, starting from \cite{BenGeloun:2011rc}, dealt with simplified TGFTs, which are going to be also our object of analysis.

The FRG provides an exact flow equation for action functionals $\Gamma_N$. It follows from the derivation of the FRG that $\Gamma_N$ possesses the field content and the symmetries that are shared by the bare action, the functional measure and the IR-suppression term. In continuum field theory one refers to the space of action functionals of a given field content and given symmetries as theory space. Continuum field theory however implies implicitly a notion of locality and dimensionality of operators which are not implicitly implied in the discrete and background independent models that we consider here. It follows that one has to specify at least a notion of dimensionality (i.e. scaling with the RG scale $N$) of operators when defining a theory space for discrete models. Since we are interested in investigating the large $N$-behavior through the FRG, we need to impose that the assignment of dimension is such that all beta functions admit a $1/N$ expansion at large $N$. Thus, given an IR-suppression term, we obtain a number of bounds for each coupling dimensionality implied by the $1/N$ expandability of the beta functions.

In practical calculations one is usually forced to make a truncation ansatz for the effective average action, i.e. one expands as
\begin{equation}
\Gamma_N = \sum_i \bar\lambda^i_N\mathcal{O}_i (\phi)\,,
\label{frg2}
\end{equation}
where $\mathcal{O}_i(\phi)$ denote contracted field operators compatible with the field content and symmetries of the underlying theory space. The $\bar\lambda^i_N$ denote the corresponding dimensionful coupling constants. The scale derivative $\partial_t \equiv N\partial_N$ acts on the truncation ansatz as
\begin{equation}
\partial_t \Gamma_N = \sum_i (\partial_t \bar\lambda^i_N)\mathcal{O}_i (\phi)\equiv\sum_i \bar\beta_i\mathcal{O}_i (\phi)\,,
\label{frg3}
\end{equation}
where $\bar{\beta}_i$ is the bare beta function of the coupling $\bar\lambda^i_N$. By evaluating the trace on the right-hand side of \eqref{frg1}, it is possible to extract the bare beta functions $\bar\beta_i$ through the application of a suitable projection of the theory space onto the truncation ansatz. 

The normalization of the field $\phi\to\sqrt{Z_N}\,\phi$ can be absorbed as a normalization of the functional measure $[d\varphi]_{N^\prime}$ as along as the UV-cutoff $N^\prime$ is finite. One can thus adopt a convention that one coupling constant $Z_N$ is redundant and associated with the normalization of the fields. One commonly chooses the quadratic kinetic term to be normalized, and associated the coupling constant $Z_N$ with the kinetic term and defines the anomalous dimension as $\eta:=\partial_t\ln(Z_N)$. We then define the dimensionless couplings by removing this redundancy and the dimensional scaling
\begin{equation}
  \bar\lambda_N^i=N^{\textrm{dim}(\mathcal O_i)}\,Z^{\Phi(\mathcal O_i)}_N\,\lambda^i_N,
\end{equation}
where $\textrm{dim}$ denotes the scaling dimension and where $2 \Phi(\mathcal O_i)$ denotes the number of fields $\phi$ that appear in the monomial $\mathcal O_i(\phi)$. We thus obtain the beta functions
\begin{equation}
  \beta_i=-\left(\Phi(\mathcal O_i)\,\eta+\textrm{dim}(\mathcal O_i)\right)\lambda^i+N^{-\textrm{dim}(\mathcal O_i)}\,Z^{-\Phi(\mathcal O_i)}_N\,\bar\beta_i.
\end{equation}
The aforementioned consistency relations for the dimensional scaling are then obtained by demanding that the leading power in $N$ of $N^{-\textrm{dim}(\mathcal O_i)}\,\bar\beta_i[N,\bar\lambda]$ is non-positive for all $i$. 

The large-$N$ limit of the FRG defines an autonomous vector field on theory space. Fundamentally, we want to investigate the possible non-trivial UV-scalings, which appear as compactly contained attractors of this autonomous vector field, in the simplest case a fixed point, i.e. a point $\lambda^i=\lambda^i_*$ where all dimensionless beta functions vanish simultaneously. Linearizing the flow near the fixed point $\lambda^i_*$ gives
\begin{equation}
  \lambda^i(N)=\lambda^i_*+\sum_I\,C_I\,V_I^i\left(\frac{N}{N_o}\right)^{-\theta_I},
\end{equation}
where $V_I^j$ denote the eigendirections $\left.\frac{\partial \beta_i}{\partial\lambda^j}\right|_{\lambda=\lambda_*}\,V^j_I=\theta_I\,V^i_I$ (here summation is implied over $j$, but not over $I$) associated with the critical exponent $\theta_I$. Positive critical exponents imply that the associated eigendirection is an IR-relevant interaction.

In practical calculations one projects the FRG vector field onto a truncation, which means that a point at which the vector fields is purely vertical (w.r.t. the projection) then this point appears as a spurious RG fixed point in the truncation. The generation of spurious fixed points by purely vertical RG flow suggests two strategies to test whether a fixed point found in a truncation is spurious: 1. one enlarges the truncation, so it becomes more likely that the flow will possess a non-vanishing horizontal component, and 2. one varies the projection rule and/or RG scheme, so the notion of ``vertical" changes. To effectively apply these two strategies for identifying true fixed points one uses two general observations: 1. {\it Universality:} quantities such as critical exponents of true fixed points change only moderately when enlarging the truncation and/or changing the RG scheme. 2. {\it Dimensionality:} If one finds a fixed point in a truncation and then enlarges the truncation by including all operators that possess scaling dimension one or two higher than the scaling dimension in the truncation and the newly introduced coupling obtain a fixed point scaling that is very close to their scaling dimension then it becomes very unlikely that the operators outside this truncation which have even higher scaling dimensions will change the fixed point behaviour. We see in particular that in order to get reliable information about a fixed point one needs to use a truncation in which all relevant directions are included.

\section{Setting the stage: the model}
\begin{figure}
	\centering
		\includegraphics[width=0.14\textwidth]{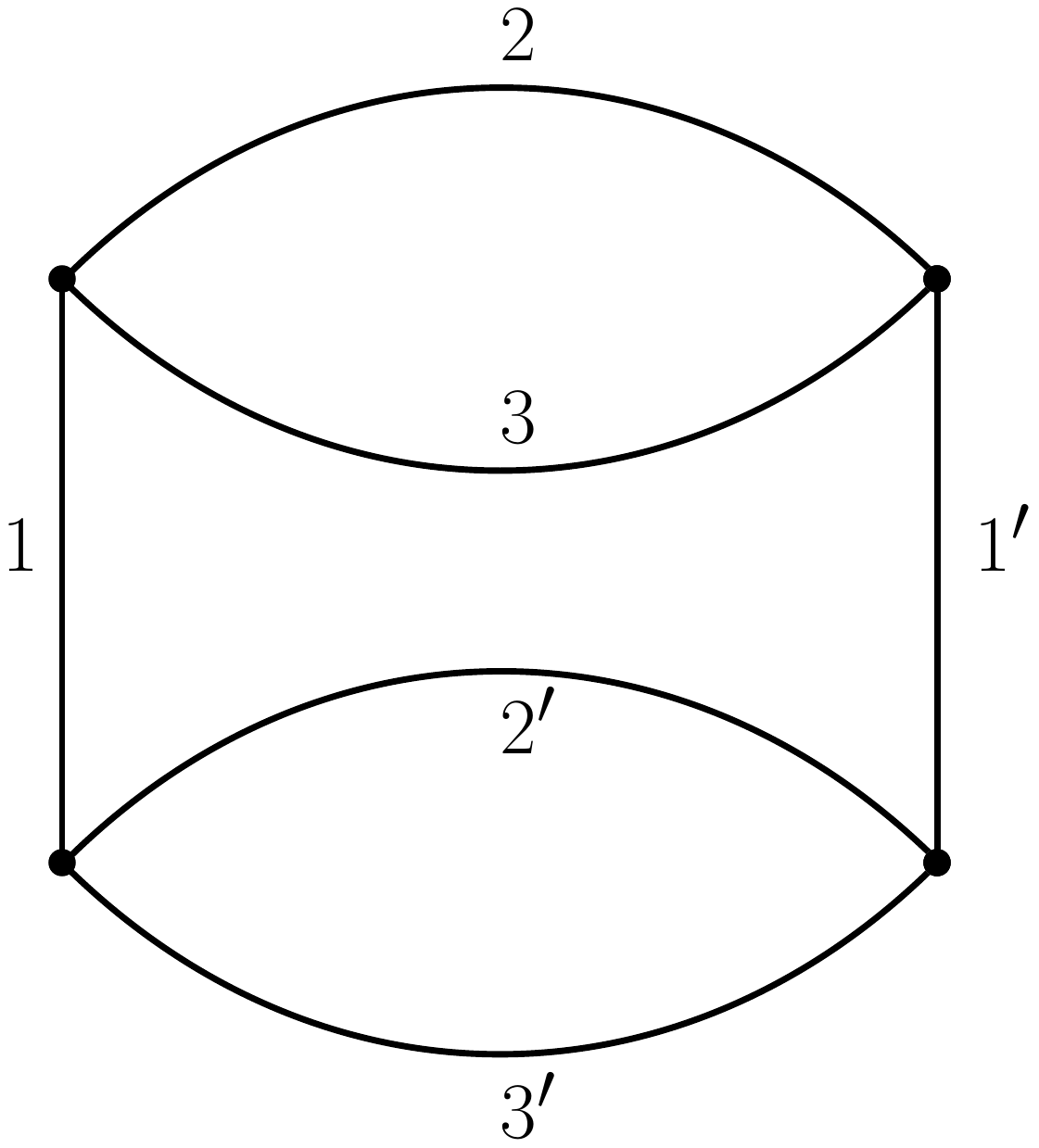}\,\,\,\,\
		\includegraphics[width=0.14\textwidth]{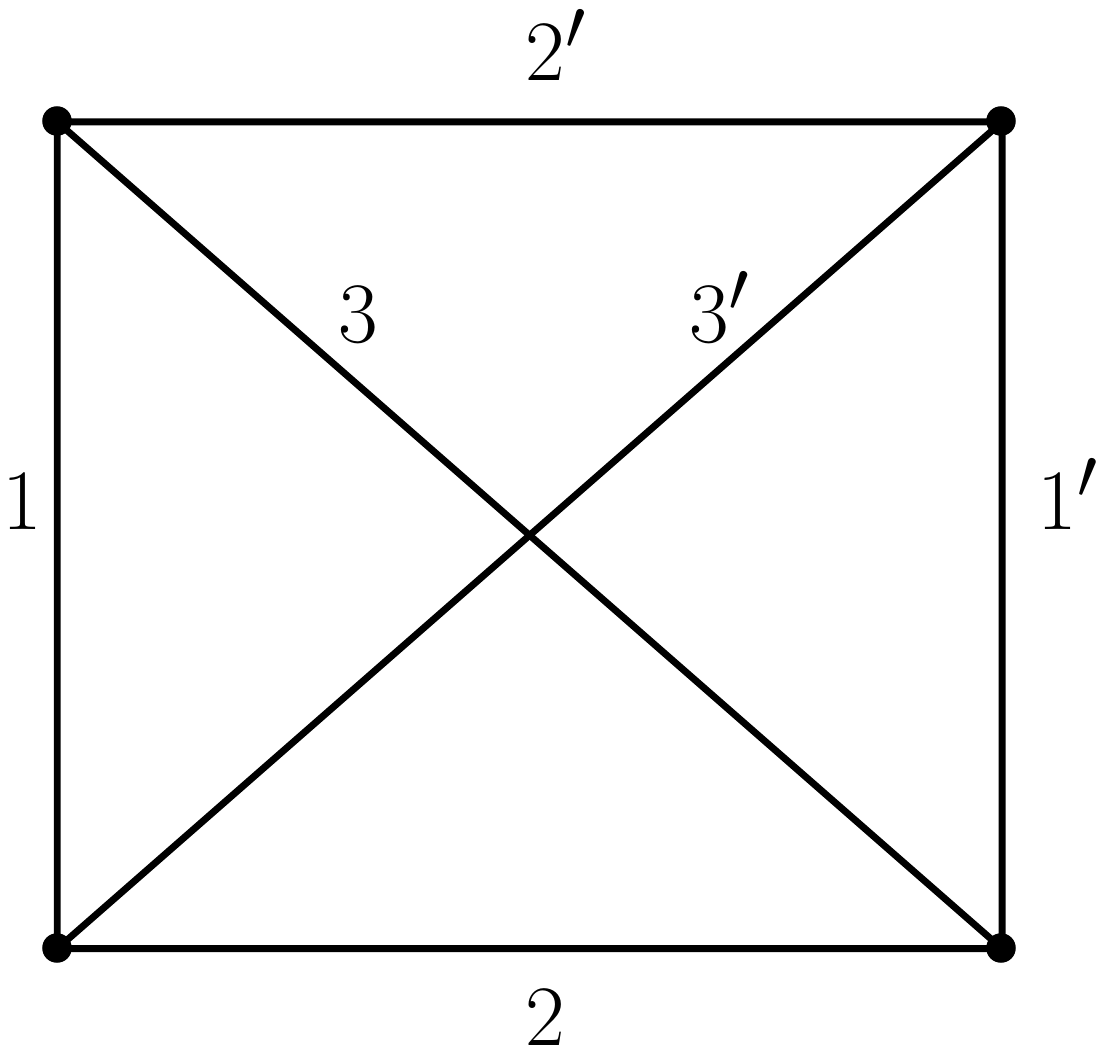}\,\,\,\,\,
		\includegraphics[width=0.14\textwidth]{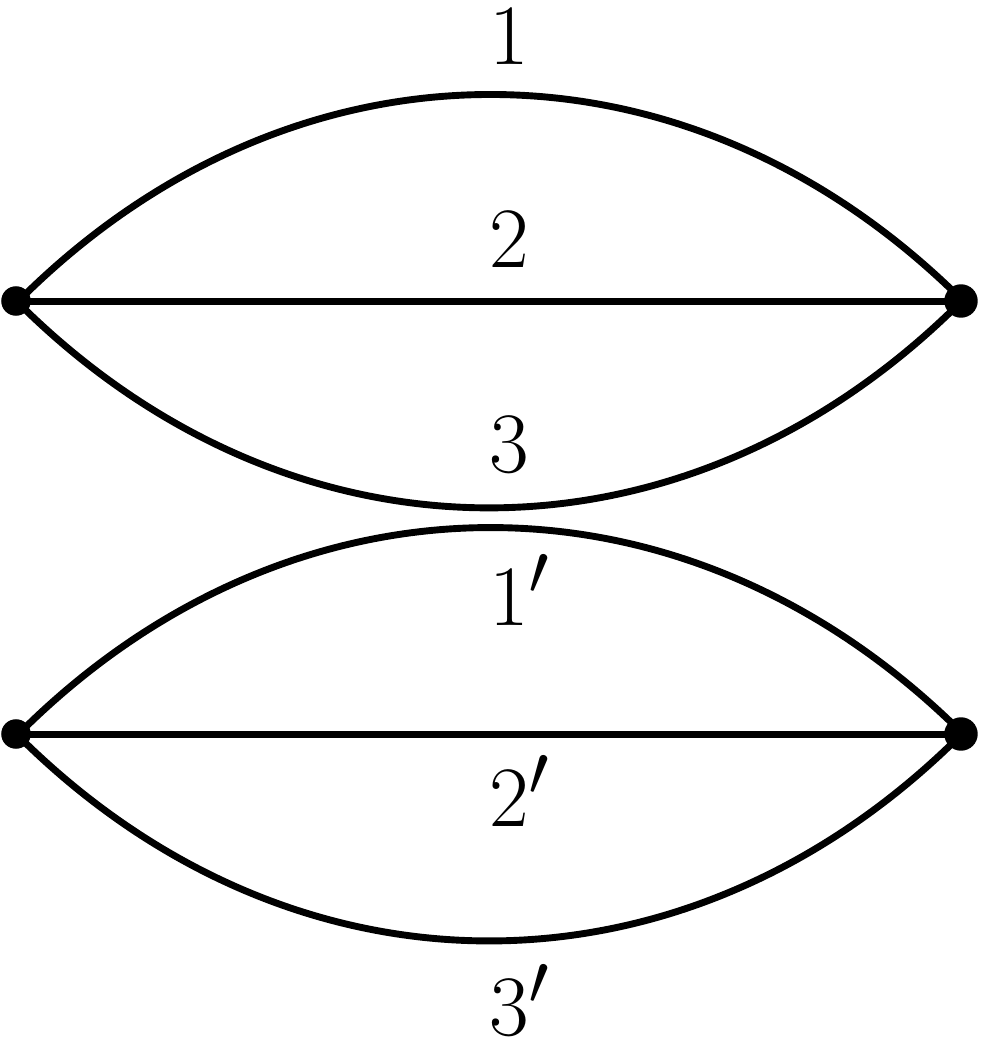}
		\caption{Interactions considered in the present truncation.}
	\label{fig:a}
\end{figure}

In this work we generalize the work of \cite{Benedetti:2014qsa} where the following truncation for the effective average action $\Gamma_N$ was considered 
\begin{equation}
\Gamma_N \left[\phi\right]=\frac{Z_N}{2}\mathrm{Tr}\left(\phi\cdot K\cdot\phi\right)+\frac{m_N}{2}\text{Tr}(\phi^2)+\Gamma^{\mathrm{int}}_N\left[\phi\right]\,,
\label{eaa1}
\end{equation}
where $Z_N$ stands for the wave-function renormalization, $m$ a mass parameter, $\phi$ is a real field over $U(1)^3$ and the kinetic kernel $K$ is defined by
\begin{equation}
K(\left\{p_i \right\};\left\{p'_i \right\}) = \delta^{(3)}_{p_i ,p'_i }\left( \frac{1}{3}\sum^ 3_{i=1}|p_i|\right)\,,
\label{eaa2}
\end{equation}
with
\begin{equation}
\delta^{(3)}_{p_i ,p'_i } = \prod^ 3_{i=1} \delta_{p_i ,p'_i }\,,
\label{eaa3}
\end{equation}
The interaction term encoded in $\Gamma^{\mathrm{int}}_N$ is written as
\begin{eqnarray}
\Gamma^{\mathrm{int}}_N&=&\frac{\lambda^{4;1}_N}{4}\sum_{p_i,p'_i\in\mathbb{Z}}\phi_{123}\phi_{12'3'}\phi_{1'2'3'}\phi_{1'23}\nonumber\\
&+&\mathrm{Sym}(1\rightarrow 2\rightarrow 3)\,.
\label{eaa4}
\end{eqnarray}
The non-local structure of the interactions on top of tensorial invariance allows the introduction of two more operators with four fields $\phi$. Hence, the term \eqref{eaa4} does not correspond to all possible interaction terms at quartic order which are momentum-independent. In this work, all quartic order and momentum-independent operators are included in the truncation, namely,
\begin{equation}
\Gamma^{\mathrm{int}}_N = \Gamma^ {4;1}_N + \Gamma^ {4;2}_N + \Gamma^ {4;3}_N\,,
\label{eaa41}
\end{equation}
with
\begin{eqnarray}
\Gamma^ {4;1}_N&=&\frac{\lambda^{4;1}_N}{4}\sum_{p_i,p'_i\in\mathbb{Z}}\phi_{123}\phi_{12'3'}\phi_{1'2'3'}\phi_{1'23}\nonumber\\
&+&\mathrm{Sym}(1\rightarrow 2\rightarrow 3)\,,\nonumber\\
\Gamma^{4;2}_N&=&\frac{\lambda^{4;2}_N}{4}\sum_{p_i,p'_i\in\mathbb{Z}}\phi_{123}\phi_{1'2'3}\phi_{1'23'}\phi_{12'3'}\,,\nonumber\\
\Gamma^{4;3}_N&=&\frac{\lambda^{4;3}_N}{4}\sum_{p_i,p'_i\in\mathbb{Z}}\phi_{123}\phi_{123}\phi_{1'2'3'}\phi_{1'2'3'}\,.
\label{eaa42}
\end{eqnarray}
It is convenient to represent these interactions using diagrams as in Fig.~\ref{fig:a}. It should be understood that the first diagram has to be symmetrized as indicated in eq.\eqref{eaa4}.
Having the ansatz for $\Gamma_N$ defined, we calculate itsRG  flow using \eqref{frg1}. For this, we have to specify a cutoff function $R_N$ and compute the Hessian $\Gamma^{(2)}_N$. The contributions to the Hessian $\Gamma^{(2)}_N$ coming from the interaction term $\Gamma^{\mathrm{int}}_N$ are
\begin{eqnarray}
F_1\left(\left\{p_i\right\};\left\{p'_i\right\}\right)&=&\lambda^{4;1}_{N}\Bigg(\phi_{p_1,p'_2,p'_3}\phi_{p'_1,p_2,p_3}\nonumber\\
&+&\sum_{m_1}\delta_{p_1,p'_1}\phi_{m_1,p'_2,p'_3}\phi_{m_1,p_2,p_3}\nonumber\\
&+&\sum_{m_2,m_3}\delta_{p_2,p'_2}\delta_{p_3,p'_3}\phi_{p'_1,m_2,m_3}\phi_{p_1,m_2,m_3}\Bigg)\nonumber\\
&+&\mathrm{Sym}(1\rightarrow 2\rightarrow 3)\,,
\label{eaas2}
\end{eqnarray}
\begin{eqnarray}
F_2\left(\left\{p_i\right\};\left\{p'_i\right\}\right)&=&\lambda^{4;2}_{N}\Bigg(\sum_{m_3}\delta_{p_3,p'_3}\phi_{p'_1,p_2,m_3}\phi_{p_1,p'_2,m_3}\nonumber\\
&+&\sum_{m_2}\delta_{p_2,p'_2}\phi_{p'_1,m_2,p_3}\phi_{p_1,m_2,p'_3}\nonumber\\
&+&\sum_{m_1}\delta_{p_1,p'_1}\phi_{m_1,p_2,p'_3}\phi_{m_1,p'_2,p_3}\Bigg)\,.
\label{eaas3}
\end{eqnarray}
\begin{eqnarray}
F_3\left(\left\{p_i\right\};\left\{p'_i\right\}\right)&=&\lambda^{4;3}_{N}\Bigg(\sum_{m_1,m_2,m_3}\delta_{p_1,p'_1}\delta_{p_2,p'_2}\delta_{p_3,p'_3}\nonumber\\
&\times&\phi_{m_1,m_2,m_3}\phi_{m_1,m_2,m_3}
+2\phi_{p_1,p_2,p_3}\nonumber\\
&\times&\phi_{p'_1,p'_2,p'_3}\Bigg)\,.
\label{eaas4}
\end{eqnarray}

The cutoff function we introduce is chosen to be the optimized one \cite{Litim:2001up},
\begin{eqnarray}
R_N (\left\{p_i\right\},\left\{p'_i\right\}) &=& Z_N \delta^{(3)}_{p_i ,p'_i }\left(N-\frac{1}{3}\sum^ 3_{i=1}|p_i|\right)\nonumber\\
&\times&\theta\left(N-\frac{1}{3}\sum^ 3_{i=1}|p_i| \right)\,.
\label{eaas5}
\end{eqnarray}
Also, it is part of the flow equation the derivative of the cutoff function with respect to $N$. This yields
\begin{eqnarray}
\partial_t R_N &=&  \delta^{(3)}_{p_i ,p'_i }\left[(\partial_t Z_N)\left(N-\frac{1}{3}\sum^ 3_{i=1}|p_i|\right)+Z_N N \right]\nonumber\\
&\times& \theta\left(N-\frac{1}{3}\sum^ 3_{i=1}|p_i| \right)\,.
\label{eaas7}
\end{eqnarray}
The contribution to the Hessian from the quadratic part of the effective average action with the addition of $R_N$ is
\begin{eqnarray}
P_N (\left\{p_i\right\},\left\{p'_i\right\}) &=& Z_N K(\left\{p_i \right\};\left\{p'_i \right\}) + m_N \delta^{(3)}_{p_i,p'_i}\nonumber\\
&+&R_N (\left\{p_i\right\},\left\{p'_i\right\})\,.
\label{eaas9}
\end{eqnarray}
Therefore, the complete Hessian is expressed as
\begin{equation}
\Gamma^{(2)}_N+R_N = P_N + F_1 + F_2 + F_3\,,
\label{eaas10}
\end{equation}
where $\Gamma^{(2)}_N$ is a short-hand notation for the second functional derivative of $\Gamma_N$. The FRG equation can be expressed in the following form
\begin{eqnarray}
\partial_t \Gamma_N &=&\frac{1}{2}\mathrm{Tr}\left[(\partial_t R_N)P^{-1}_N
+\sum^{\infty}_{n=1}(-1)^ n (\partial_t R_N)P^{-1}_N\right.\nonumber\\
&\times&\left.\left(F_1 P^{-1}_N+F_2 P^{-1}_N+F_3 P^{-1}_N\right)^ n\right]\,.
\label{eaas11}
\end{eqnarray}
The first term of the right-hand side of \eqref{eaas11} is field independent and thus corresponds to a vacuum term which can be absorbed into the normalization of the functional measure, so we can discard it for our considerations. We start by looking to terms quadratic in the fields which corresponds to the contributions for $n=1$. For such a contribution, one has
\begin{equation}
\partial_t \Gamma_N \Big|_{n=1} = -\frac{1}{2}\mathrm{Tr}\left[(\partial_t R_N)P^{-1}_N (F_1+F_2+F_3)P^{-1}_N\right]\,.
\label{eaas12}
\end{equation}
Since our truncation involves up to quartic order terms on the fields, we have to consider up to $n=2$ contributions. These can be written as
\begin{eqnarray}
\partial_t \Gamma_N \Big|_{n=2}&=&\frac{1}{2}\mathrm{Tr}\left[(P^{-1}_N (\partial_t R_N) P^{-1}_N)_{p_i}(F_1+F_2+F_3)_{p_i,p'_i}\right.\nonumber\\
&\times&\left.P^{-1}_N (\left\{p'_i\right\},\left\{p''_i\right\})(F_1+F_2+F_3)_{p''_i,p_i}\right]\,.
\label{eaas13}
\end{eqnarray}
Also, we use a condensed notation whenever is possible. 

For $n>2$, one gets terms which have six or more fields. Hence, those terms do not belong to the proposed truncation \eqref{eaa1} and we disregard them. Having \eqref{eaas11}, \eqref{eaas12} and \eqref{eaas13} we can perform explicit computations. 

In \cite{Benedetti:2014qsa} the explicit computations in the absence of the $F_2$ and $F_3$ contributions were presented in details. The extension of the computation to $F_2$ and $F_3$ is straightforward (although lenghty) and brings no technical novelty that needs to be stressed. Hence, for simplicity, we do not report the details here. As discussed in \cite{Benedetti:2014qsa}, the system of equations derived from the FRG equation is non-autonomous for a generic value of $N$. In particular the flow equation leads to
\begin{widetext}
\begin{equation}
\eta = \frac{18N\left(2\lambda^{4;2}_{N}+3\lambda^{4;1}_{N} (2N+1)\right)}{3\left((m_{N}+N)^2-2\lambda^{4;3}_{N}-12\lambda^{4;2}_{N}N\right)-2 \lambda^{4;1}_{N} \left(27N^2+18N+5\right)}\,,
\label{beta1}
\end{equation}
\begin{eqnarray}
\partial_{t}m_{N}&=&-\frac{N}{(m_{N}+N)^2}\left[\left(9\lambda^{4;1}_{N} \left(6N^2+4N+1\right)+\lambda^{4;3}_{N} \left(36N^3+18N^2+8N+3\right)\right.\right.\nonumber\\
&+&\left.\left.3\lambda^{4;2}_{N} (6N+1)\right)+\eta \left(\lambda^{4;3}_{N} \left(9N^3+2N+2\right)+\lambda^{4;1}_{N} \left(18N^2+9N+4\right)+9\lambda^{4;2}_{N} N\right)\right]\nonumber\\
&-&\eta m_{N}\,,
\label{beta2}
\end{eqnarray}
\begin{eqnarray}
\partial_{t}\lambda^{4;1}_{N}&=& \frac{2N}{(m_N+N)^3}\left[3\left(6 (\lambda^{4;1}_{N})^2 (N+1)^2+4\lambda^{4;1}_{N} \left(\lambda^{4;3}_{N}+\lambda^{4;2}_{N}(2N+1)\right)+(\lambda^{4;2}_{N})^2 (2N+1)\right)\right.\nonumber\\
&+&\left.\frac{\eta}{3}\left((\lambda^{4;1}_{N})^2 \left(18 N^2+45N+37\right)+12 \lambda^{4;1}_{N} \left(3\lambda^{4;3}_{N}+\lambda^{4;2}_{N}(3N+2)\right)+3(\lambda^{4;2}_{N})^2 (3N+2)\right)\right]\nonumber\\
&-&2\lambda^{4;1}_{N}\eta\,,
\label{beta3}
\end{eqnarray}
\begin{eqnarray}
\partial_{t}\lambda^{4;2}_N &=& \frac{12N}{(m_N+N)^3}\left[2\left((\lambda^{4;1}_{N})^2+\lambda^{4;2}_{N} \lambda^{4;3}_{N}+(3N+1)\lambda^{4;2}_{N} \lambda^{4;1}_{N}\right)+ \eta\left(2(\lambda^{4;1}_{N})^2+2 \lambda^{4;2}_{N} \lambda^{4;3}_{N}\right.\right.\nonumber\\
&+&\left.\left.(3N+1)\lambda^{4;2}_{N} \lambda^{4;1}_{N}\right)\right]-2\lambda^{4;2}_{N}\eta\,,
\label{beta4}
\end{eqnarray}
\begin{eqnarray}
\partial_{t}\lambda^{4;3}_{N}&=&\frac{2N}{(m_N+N)^3}\left[ \left(6\lambda^{4;1}_{N}\left(\lambda^{4;2}_{N}+3\lambda^{4;3}_{N}\left(6N^2+4N+1\right)\right)+\lambda^{4;3}_{N}\left(\lambda^{4;3}_{N} \left(36N^3+18N^2+8N+9\right)\right.\right.\right.\nonumber\\
&+&\left.\left.\left.6\lambda^{4;2}_{N}(6N+1)\right)+3(\lambda^{4;1}_{N})^2(12N+5)\right)+\eta\left(\lambda^{4;3}_{N} \left(\lambda^{4;3}_{N} \left(9N^3+2N+8\right)+18\lambda^{4;2}_{N}N\right)\right.\right.\nonumber\\
&+&\left.\left.2\lambda^{4;1}_{N}\left(3 \lambda^{4;2}_{N}+\lambda^{4;3}_{N} \left(18N^2+9N+4\right)\right)+9(\lambda^{4;1}_{N})^2 (2N+1)\right)\right]-2\lambda^{4;3}_{N}\eta\,,
\label{beta5}
\end{eqnarray}
\end{widetext}
where we have performed the redefinitions
\begin{equation}
(m_N,\lambda^{3;i}_N)\,\to\,(Z_N m_N,Z^2_N \lambda^{4;i}_N)\,,
\label{redef1}
\end{equation}
and $\eta = \partial_t \mathrm{ln}Z_N$ is the anomalous dimension. Clearly, the beta functions described by eqs.\eqref{beta1} - \eqref{beta5} form a non-autonomous system due to the explicit dependence on $N$. Nevertheless, these equations are written as flow equations for the ``dimensionful" couplings $(m_N,\lambda^{4;i}_N)$. As usual, one should rescale the couplings in such a way that the flow equations are expressed in terms of the dimensionless couplings. In ordinary quantum field theories, the beta functions for dimensionless couplings form an autonomous system unless some external scale is present in the theory in such a way that dimensionless ratios can be constructed. In the present case, there is no possible redefinition of the couplings which turns eqs.\eqref{beta1} - \eqref{beta5} in an autonomous system. The reason behind this fact was explored in details in \cite{Benedetti:2014qsa,Benedetti:2015yaa} and is related to the existence of an external scale in the theory, namely, the radius of the (compact) group manifold. For non-compact groups, one sees that the beta functions are autonomous, see \cite{Geloun:2015qfa,Geloun:2016qyb}. Therefore the analysis of fixed-points for a generic $N$ becomes extremely difficult. Nevertheless, it is possible to obtain autonomous beta functions systems for $N\gg 1$, i.e. the deep UV. In the following sections, we perfom the analysis in different truncations.

\section{Large $N$ limit: without double-trace interactions}

In this section we restrict the analysis to the case where $\lambda^{4;3}_N =0$. In the large $N$ limit, we have to rescale the couplings in such a way that the dependence on $N$ is canceled out on the beta functions. The criterium to obtain a well-defined (and non-trivial) autonomous beta-functions system leads to the following redefinitions of the couplings
\begin{equation}
m_N \to  N \bar{m}_N\,,\,\, \lambda^{4;1}_N \to \bar{\lambda}^{4;1}_N\,,\,\, \lambda^{4;2}_N\to \bar{\lambda}^{4;2}_N N^\alpha\,,
\label{lnlm1}
\end{equation}
with $-2 \leq \alpha \leq 1/2$. It is not possible to determine uniquely the value for $\alpha$. From the functional renormalization point of view, this ambiguity is potentially fixed by considering more sophisticated truncations, see also \cite{Eichhorn:2017xhy}. The structure of the beta functions changes if the upper bound for $\alpha$ is saturated or not. For clarity reasons, we separate the discussion for these two cases. Therefore, at the large $N$ limit, the beta function system is expressed, for $\alpha <1/2$, as
\begin{eqnarray}
\eta&=&\frac{36\bar{\lambda}^{4;1}_{N}}{(1 + \bar{m}_{N})^2-18\bar{\lambda}^{4;1}_{N}}\,,\nonumber\\
\partial_{t}\bar{m}_{N}&=&-(1+\eta)\bar{m}_{N}-18(3+\eta)\frac{\bar{\lambda}^{4;1}_N}{(1+\bar{m}_{N})^2}\,,\nonumber\\
\partial_{t}\bar{\lambda}^{4;1}_{N}&=&12(3+\eta)\frac{(\bar{\lambda}^{4;1}_{N})^2}{(1+\bar{m}_{N})^3}-2\eta\bar{\lambda}^{4;1}_{N}\,,\nonumber\\
\partial_{t}\bar{\lambda}^{4;2}_{N}&=&-(2\eta+\alpha)\bar{\lambda}^{4;2}_{N}\,.\nonumber\\
\label{lnlm2}
\end{eqnarray}
Clearly, the non-melonic coupling $\bar{\lambda}^{4;2}_{N}$ decouples from the rest of the beta functions. At this stage, it is easy to compute the fixed-points from eq.\eqref{lnlm2}. In fact, in this case, besides the Gaussian fixed point $\bar{m}_{\ast}=\bar{\lambda}^{4;1}_\ast=\bar{\lambda}^{4;2}_\ast=0$ one gets
\begin{align}
\bar{m}_{\ast} &= -0.7926\,, &&\bar{\lambda}^{4;1}_{\ast}=0.0042\,, &&&\bar{\lambda}^{4;2}_{\ast}=0\,, \label{fpnm1}\\
\bar{m}_{\ast} &= -0.5407\,, &&\bar{\lambda}^{4;1}_{\ast}= 0.0028\,, &&&\bar{\lambda}^{4;2}_{\ast} = 0\,, \label{fpnm2}
\end{align}
which are independent of $\alpha$. The fixed points agree with  \cite{Benedetti:2014qsa}. In particular, we emphasize that the fixed point defined by eq.\eqref{fpnm1} is discarded due to the fact that it cannot be connected to the Gaussian due to the existence of a singularity, see  \cite{Benedetti:2014qsa}.  Also, the computations of the critical exponents associated to \eqref{fpnm1} lead to huge critical exponents, suggesting that this fixed point is a truncation artifact. For the fixed point \eqref{fpnm2}, the critical exponents are
\begin{equation}
\theta = \left\{-1.8682,0.8571,1.2915+\alpha\right\}\,.
\label{ce1}
\end{equation}
One notices that albeit the fixed point value \eqref{fpnm2} is $\alpha$-independent, the critical exponents are not. From \eqref{ce1} it is clear that the number of relevant directions depends on the value of $\alpha$. This ambiguity shows up because the functional renormalization group is not able to fix uniquely the scaling dimension for the coupling $\lambda^{4;2}_N$ associated with the non-melonic vertex. Clearly, if $\alpha < -1.2915$ the fixed point has just one relevant direction. However, for the allowed range of values of $\alpha$, it is also possible that such a fixed point has two relevant directions. A definite answer requires improvement of the present truncation by including vertices with higher power of fields \textit{e.g.} $\phi^6$ interactions. This is the first subtlety we report in this paper regarding the enlargement of the truncation employed in \cite{Benedetti:2014qsa}.

For completeness, let us consider the case $\alpha=1/2$ namely, when it saturates the upper bound allowed by the present truncation. The anomalous dimension and the beta function for the mass are the same as those written in \eqref{lnlm2}. For the couplings $(\bar{\lambda}^{4;1}_N,\bar{\lambda}^{4;2}_N)$, the beta functions for $\alpha = 1/2$ are
\begin{eqnarray}
\partial_{t}\bar{\lambda}^{4;1}_{N}&=&12(3+\eta)\frac{(\bar{\lambda}^{4;1}_{N})^2}{(1+\bar{m}_{N})^3}-2\eta\bar{\lambda}^{4;1}_{N}\nonumber\\
&+& 6(2+\eta)\frac{(\bar{\lambda}^{4;2}_N )^2}{(1+\bar{m}_N)^3}\,,\nonumber\\
\partial_{t}\bar{\lambda}^{4;2}_{N}&=&-\left(2\eta+\frac{1}{2}\right)\bar{\lambda}^{4;2}_{N}\,.\nonumber\\
\label{alpha1/2}
\end{eqnarray}
The coupling associated with the non-melonic interaction $\bar{\lambda}^{4;2}_{N}$ appears in the beta function of $\lambda^{4;1}_N$ associated to the melonic interaction. This is different from the previous situation ($\alpha <1/2)$ where $\bar{\lambda}^{4;2}_N$ decouples from the beta function of the other couplings. In this case, we obviously get the same fixed points \eqref{fpnm1} and \eqref{fpnm2} next to 
\begin{equation}
\bar{m}_{\ast} = 0.5238\,, \qquad\bar{\lambda}^{4;1}_{\ast}=-0.0184\,, \qquad\bar{\lambda}^{4;2}_{\ast}=\pm 0.0451\,, \label{fpnm3}
\end{equation}
which amounts to set $\eta=-1/4$. The critical exponents associated with \eqref{fpnm3} are
\begin{equation}
\theta_{1,2} = \left\{1.3892,-0.3313-0.2126 i, -0.3313 + 0.2126i\right\}\,.
\label{critexpo2}
\end{equation}
From \eqref{ce1} and \eqref{critexpo2} one easily sees that all the fixed points for $\alpha=1/2$ have two relevant directions. We must emphasize that $\alpha=1/2$ is the upper bound that the present truncation allows for $\alpha$. Also, within the allowed range for $\alpha$, this is the only value which results on the appearance of the non-melonic coupling in beta functions for different couplings than $\bar{\lambda}^{4;2}_N$. The only way, within the functional renormalization group framework, to fix the value of $\alpha$ is by considering further terms in the truncation for the effective average action. A similar feature was observed in \cite{Eichhorn:2017xhy} in the context of pure tensor models. Still in the realm of pure tensor models, it is possible to show that by enlarging the truncation from quartic order vertices to sixth ones, this upper bound for $\alpha$ is lowered, see \cite{EKLP}.  

\section{Large $N$ limit: with double-trace interactions}

In this case, we keep the coupling associated to the double-trace vertex $\lambda^{4;3}_N$. Taking the large $N$ limit, the beta functions display a well-defined and non-trivial large $N$ expansion if the couplings $(m_N,\lambda^{4,i}_N)$ with $i=1,2,3$ are rescaled as eq.\eqref{lnlm1} on top of
\begin{equation}
\lambda^{4;3}_N = \bar{\lambda}^{4;3}_N N^{-1}\,.
\label{dti1}
\end{equation}
Such a rescaling is uniquely determined by the system of beta functions within the present truncation. Again, we perform the analysis for $\alpha < 1/2$ and $\alpha = 1/2$. Hence, the beta functions system for $\alpha < 1/2$ at large $N$ is expressed as
\begin{eqnarray}
\eta&=&\frac{36\bar{\lambda}^{4;1}_{N}}{(1 + \bar{m}_{N})^2-18\bar{\lambda}^{4;1}_{N}}\,,\nonumber\\
\partial_{t}\bar{m}_{N}&=&-(1+\eta)\bar{m}_{N}-18(3+\eta)\frac{\bar{\lambda}^{4;1}_N}{(1+\bar{m}_{N})^2}\nonumber\\
&-&9(4+\eta)\frac{\bar{\lambda}^{4;3}_N}{(1+\bar{m}_N)^2}\,,\nonumber\\
\partial_{t}\bar{\lambda}^{4;1}_{N}&=&12(3+\eta)\frac{(\bar{\lambda}^{4;1}_{N})^2}{(1+\bar{m}_{N})^3}-2\eta\bar{\lambda}^{4;1}_{N}\,,\nonumber\\
\partial_{t}\bar{\lambda}^{4;2}_{N}&=&-\left(2\eta+\alpha\right)\bar{\lambda}^{4;2}_{N}\,,\nonumber\\
\partial_t \bar{\lambda}^{4;3}_{N}&=& 36(2+\eta)\frac{(\bar{\lambda}^{4;1}_N)^2}{(1+\bar{m}_N)^3} +72(3+\eta)\frac{\bar{\lambda}^{4;1}_N \bar{\lambda}^{4;3}_N}{(1+\bar{m}_N)^3}\nonumber\\
&+&18(4+\eta)\frac{(\bar{\lambda}^{4;3}_N)^2}{(1+\bar{m}_N)^3}-(2\eta-1)\bar{\lambda}^{4;3}_N\,.
\label{dti2}
\end{eqnarray}

\begin{table}[tbp]
\centering
\label{TableI}
\begin{tabular}{|c|c|c|c|}
\hline
$\bar{m}_\ast$ & $\bar{\lambda}^{4;1}_\ast$ & $\bar{\lambda}^{4;2}_\ast$ & $\bar{\lambda}^{4;3}_\ast$ \\ \hline
$-0.3990$      & $-0.0122$                  & $0$                        & $0.0180$                   \\ \hline
$1.0000$       & $0$                        & $0$                        & $-0.1111$                  \\ \hline
\end{tabular}
\caption{Fixed points for $\alpha < 1/2$.}
\end{table}

\begin{table}[tbp]
\centering
\label{TableII}
\begin{tabular}{|c|c|c|c|}
\hline
$\theta_1$          & $\theta_2$          & $\theta_3$ & $\theta_4$ \\ \hline
$-1.5528-2.1323\,i$ & $-1.5528+2.1323\,i$ & $-1.5094$  & $-0.7506$  \\ \hline
$2.8229$            & $0.1771$            & $0$        & $0$        \\ \hline
\end{tabular}
\caption{Critical exponents for the correspondent fixed points in Table I for $\alpha=0$.}
\end{table}

The fixed points obtained from eqs.\eqref{dti2} are displayed in Table~I. We should emphasize they are independent of the parameter $\alpha$ and for all of them, the non-melonic coupling fixed point $\bar{\lambda}^{4;2}_\ast$ vanishes. On the other hand, the fixed points for the coupling associated to the double-trace vertex $\bar{\lambda}^{4;3}_\ast$ are non-zero. Also, we have excluded fixed points for which the couplings assume complex values. Another important feature is that all the fixed points display more than one relevant direction as illustrated on Table~II (the explicit values are computed for $\alpha=0$, but they don't change qualitatively for the first fixed point while for the second fixed point, for $-2\le \alpha < 1/2$, one obtains $\theta = \left\{2.8229\,,0.1771\,,0\,,\alpha \right\}$). The first fixed point has complex critical exponents (differently from the real ones obtained in the previous section) and has a fixed point value for $\bar{\lambda}^{4;1}_\ast$ which is negative in contrast to the value obtained in eq.\eqref{fpnm2}. Most importantly, it has four irrelevant directions for $-2\leq \alpha < 1/2$. Therefore, upon the introduction of the double-trace operator, we observe a qualitative difference with respect to the previous truncation. In particular, a reduction on the number of relevant directions. Since multi-trace operators are generated by the flow equation and this simple extension discussed above indicates sharp differences between the results with and without double-trace operators one can interpret this fact as a hint that results obtained with truncations that do not take them into account can be unstable under introduction of those operators. Finally, the second fixed point is generated due to the presence of $\bar{\lambda}^{4;3}_N$ in the beta function for the mass $\bar{m}_N$. This fixed point belongs to a different class than those obtained in the truncation without the double-trace vertex.   

We report the results for $\alpha=1/2$. In this case, the coupling $\bar{\lambda}^{4;2}_N$ enters the beta function of $\bar{\lambda}^{4;1}_N$. The beta functions system is the same as the one in eq.\eqref{dti2} with the following replacements,
\begin{eqnarray}
\partial_{t}\bar{\lambda}^{4;1}_{N}&=&12(3+\eta)\frac{(\bar{\lambda}^{4;1}_{N})^2}{(1+\bar{m}_{N})^3}+6(2+\eta)\frac{(\bar{\lambda}^{4;2}_N)^2}{(1+\bar{m}_N)^3}\nonumber\\
&-&2\eta\bar{\lambda}^{4;1}_{N}\nonumber\\
\partial_{t}\bar{\lambda}^{4;2}_{N}&=&-\left(2\eta+\frac{1}{2}\right)\bar{\lambda}^{4;2}_{N}\,.
\label{dti3}
\end{eqnarray}
Besides the fixed points reported in Table~I, we obtain
\begin{eqnarray}
\bar{m}_\ast &=& 0.7738\,\qquad\bar{\lambda}^{4;1}_\ast=-0.0250\,,\qquad \bar{\lambda}^{4;2}_\ast = \pm 0.0684\,,\nonumber\\
\bar{\lambda}^{4;3}_\ast&=&-0.0175\,. 
\label{Table III}
\end{eqnarray}
with critical exponents
\begin{eqnarray}
\theta_{1,2}&=&\left\{
1.7521\,,   -0.4648-0.3282\,i\,, -0.4648+0.3282\,i\,, \right.\nonumber\\
&-&\left.0.1641\right\}\,.
\label{Table IV}
\end{eqnarray}
As it is clear from eq.\eqref{Table IV}, the fixed points reported in eq.\eqref{Table III} display just one relevant direction. Also, those fixed points have non-trivial values for all couplings. It is important to mention that the fixed points given by \eqref{Table III} look very similar to those in eq.\eqref{fpnm3} where the double-trace operator is not present. Moreover, the critical exponents in eq.\eqref{Table IV} are also rather close to \eqref{critexpo2}. The new direction that arises due to the introduction of the double-trace interaction is irrelevant. Hence, we see that when the upper-bound for the scaling dimension of the non-melonic coupling $\lambda^{4;2}_N$ is satured, the results obtained with or without the double-trace operators are similar. On the other hand, we emphasize that they are qualitatively different from the fixed point obtained in the truncation implemented in \cite{Benedetti:2014qsa}. In particular, the melonic coupling fixed point $\bar{\lambda}^{4;1}_\ast$ and the mass value $\bar{m}_\ast$ have opposite sign with respect to \cite{Benedetti:2014qsa}. However, as already pointed out, $\alpha=1/2$ is the upper bound allowed by a well-defined large $N$ expansion of the beta functions. Preliminary studies of larger truncations, namely, those involving vertices with six fields, indicate that this upper bound is lowered and, therefore, these fixed points are, potentially, just artifacts of the truncation up to quartic vertices. We will report on that elsewhere. In any case, within the present truncation, these fixed points cannot be excluded.  

\section{Comparison with pure tensor models}

One commonly expects that enlargement of the truncation leads to a better approximation of the true physical behavior, so including disconnected invariants should improve the estimates for critical behavior in the present model. The rule of thumb that a larger truncation leads to a better approximation is however not always satisfied. Here we present evidence that this is the case when including disconnected invariants in a small truncation for tensor models by considering related models, in particular pure tensor models. The necessary beta functions for the pure complex tensor model have been calculated in \cite{Eichhorn:2017xhy} (equations 24 to 35), but for the present discussion we need only: 
\begin{equation}
  \eta=\frac{5\left(g^2_{4,2}+\sum_{i=1}^3 g_{4,1}^{2,i}\right)}{20+g^2_{4,2}+\sum_{i=1}^3 g_{4,1}^{2,i}}
\end{equation}
\begin{equation}
  \beta^{2,i}_{4,1}=(2+2\eta)g^{2,i}_{4,1}+\frac{13}{630}(21-4\eta)\left(g^{2,i}_{4,1}\right)^2
\end{equation}
\begin{equation}
  \begin{array}{rcl}\beta^2_{4,2}&=&(3+2\eta)g^2_{4,2}+\frac{6-\eta}{15}\left[\left(g^2_{4,2}\right)^2+2g^2_{4,2}\sum_{i=1}^3g^{2,i}_{4,1}\right.\\
  &&\quad\quad\quad\quad\quad\quad\quad\quad\quad\left.+2\sum_{i<j=1}^3g_{4,1}^{2,i}g_{4,1}^{2,j}\right]
  \end{array}
\end{equation}
where $g^{2,i}_{4,1}$ denotes the coupling constant for the cyclic melon with preferred color $i$ and $g^2_{4,2}$ the coupling of the mass-squared term. This system of beta functions contains a number of unreasonable fixed points (e.g. with complex couplings or with unreasonably large critical exponents) and here we will only discuss those that are relevant for the comparison with the model investigated in this paper. The reasonable non-Gaussian fixed points are:
\begin{enumerate}
  \item The color symmetric NGFP at which all $g^{2,i}_{4,1}$ take the same non-vanishing value appears only in a truncation that does not include $g^2_{4,2}$. This fixed point appears at $g_{4,1}^{2}=-0.897$ with only one relevant direction with critical exponent $\theta=-2.31$. However, once $g^2_{4,2}$ is included in the truncation, one obtains only color symmetric fixed points with complex coupling constants or unreasonably large critical exponents.
  \item The one-color NGFP at which all except one $g^{2,i}_{4,1}$ vanish. It then follows from the structure for the beta function of $g^2_{4,2}$ that one can set $g^2_{4,2}=0$ at this fixed point. The nonvanishing acquires the fixed point value $g^{2,i}_{4,1}=-1.94$ and we find the critical exponents $(-2.21,0.24,0.93,0.93)$.
  \item The multi-trace NGFP at which all $g^{2,i}_{4,1}$ vanish, so only $g^2_{4,2}$ acquires a non-vanishing fixed point value at $-2.88$. At this fixed point we find the critical exponents $(-3.47,0.32,0.32,0.32)$.
\end{enumerate}
We see that the fixed points of the pure complex model have a similar structure to the ones that we found in the present paper. This is not surprising since the algebraic structure of the beta functions is dictated by the combinatorics of tensor invariants. The details of the model change only the numerical factors, which also depend on the RG scheme. Hence, one expects that physical fixed points, which exist independently of the RG scheme, appear in a wide variety of tensor models with similar characteristics. 

Using this correspondence between distinct tensor models, we see that the color symmetric and the multi-trace fixed points are the ones related to the fixed points discussed in the previous sections. Moreover, we see that the color-symmetric fixed point exhibits a similar behavior under truncation enlargement as we found on the model investigated in the present paper: It appears as a reasonable fixed point in a truncation that does not include $g^2_{4,2}$, but there seems to be no reliable (i.e. with real couplings and reasonable critical exponents) fixed point in a truncation that includes $g^2_{4,2}$.

\section{Strategies for future work}

The results of previous sections suggest that the color symmetric fixed points found in truncations that do not include multi-trace invariants for the effective average action of tensorial field theories (and of pure tensor theories) are presumably truncation artifacts. Unfortunately, the presently considered truncations are still too small to decisively distinguish truncation artifacts from physical RG fixed points, because enlargements by a single operator of a very small truncation can have drastic effects on the critical exponents of physical fixed points, such that one might falsely discard them as artifacts. To distinguish the two cases needs to apply the general criteria mentioned in section II, i.e. truncation enlargement and the study of scheme dependence. 

As a rule of thumb, one expects that one obtains reliable information about a fixed point at which only invariants with four tensors obtain essentially non-vanishing fixed point values, i.e. where the fixed point values of the remaining couplings have a scheme dependence that is about as big as their fixed point values, in truncations that include eight tensors, such that any truncation effects can at most be transmitted through ``two-loop" effects. It is therefore important to push the FRG investigation of tensorial field theories to this order and to study scheme dependence in this truncation. 

In the near future it will only be practically possible to consider truncations to eighths order with index-independent operators, thus neglecting the effect of index-dependent operators. The effect of these operators can be estimated by the $(O(N))^3$-Ward-identity, i.e. the symmetry under $\phi_{abc}\to (O_1)^d_a(O_2)^e_b(O_3)^f_c\phi_{def}$. In a pure tensor model one obtains a Ward-identity that is only broken by the presence of the regulator, while in a tensor field theory one obtains a Ward-identity that is broken by the regulator and the kinetic term. Thus, in either case one obtains a scheme dependent Ward-identity. The true critical exponents are of course scheme independent. Hence one expects that one can already obtain reasonably good results in an index-independent truncation, when one optimizes the regulator to minimize the scheme dependence. It therefore seems important to not only enlarge the truncation, but also to study the scheme dependence in these truncations in order to obtain reliable results.

Another aspect to be held in mind in the investigation of tensorial models is the universality across various models, in particular tensor models and tensorial field theories, that we used in the previous section. This is based on the observation that the models are graphically very similar, which means that the beta functions of these models become algebraically similar and hence the mechanisms due to which certain fixed points exist is similar in these models. This means that one should first enlarge the truncations in pure tensor models, where the calculations simplify, and use the lessons learned in these models to enlarge the truncation in tensorial field theories.

\section{Conclusions}

In this paper we considered the full index-independent $\phi^4$-truncation of a real $U(1)$-tensorial field theory with linear kinetic term and $(U(1))^3$-symmetry\footnote{This symmetry implies that the first indices are only contracted with first indices, second with second and third with third. However, unlike the $(O(N))^3$-symmetry, which is explicitly broken in the present model, it does allow for index-dependent terms in the effective action.}. This truncation enlarges the analysis of \cite{Benedetti:2014qsa} by including two new types of operators: 1. the truncation includes a crossed interaction term (in the notation of the present paper associated with the coupling constant $\lambda^{4;2}$), which is not a melonic interaction term; and 2. the truncations includes the double-trace interaction (in the notation of the present paper $\lambda^{4;3}$), which is a so-called disconnected interaction term. The main results of investigating this truncation are the following:
\begin{enumerate}
\item The $1/N$ expandability of the beta functions does not uniquely determine the scaling dimension of the non-melonic interaction term. Instead, one finds only non-coinciding upper and lower bounds on the scaling dimension of the coupling constant $\lambda^{4;2}$. 
\item The structure of the beta functions changes when the upper bound is saturated: For values below the upper bound, one finds that the coupling constant $\lambda^{4;2}$ does not appear in the beta functions of any other coupling, while it does appear in the beta functions of other couplings when the upper bound is saturated.
\item The number of relevant directions changes within the range given by the lower and upper bound. This means that we can not determine the universality class of the found fixed points within the present truncation.
\item We argue that the non-coincidence of the lower and upper bound is a consequence of the truncation: The present truncation seems to not yet include enough operators that effectively interact with the non-melonic sector and hence do not yet imply coinciding upper and lower bounds.
\item We find the non-Gaussian fixed point that had been found in \cite{Benedetti:2014qsa}. However, when including the double trace operator we find a significant change if the critical behavior. We argue that this behavior under truncation enlargement is not unique to this particular model. In fact, by comparing the present results with the results of the pure tensor model investigated in \cite{Eichhorn:2017xhy}, we find qualitatively the same behavior. These results suggest that one needs to consider truncations with $\phi^6$- and $\phi^8$-interactions to determine the critical behavior at this fixed point.
\end{enumerate}
The present analysis is a clear indication that the application of the FRG to tensor models and tensorial field theories has to be pushed to larger truncations, both in pure tensor models and in tensorial field theories. We expect that this enlargement of truncations will allow us to determine unique scaling dimensions for all operators that are not too close to the boundary of the truncation. Furthermore, we argued that these truncation enlargements and the study of scheme dependence in the enlarged truncations will likely resolve the questions raised in the present work. 

\section*{Acknowledgments}
We thank Astrid Eichhorn and Johannes Lumma for discussions. ADP acknowledges funding by the DFG, Grant Ei/1037-1. TK acknowledges support from the foundational questions institute through FQXi-MGA-1713 and through the PAPIIT program at UNAM through project IA103718.

\end{document}